\documentclass[sigchi]{acmart}
\usepackage{graphicx}
\AtBeginDocument{%
  \providecommand\BibTeX{{%
    \normalfont B\kern-0.5em{\scshape i\kern-0.25em b}\kern-0.8em\TeX}}}

\setcopyright{rightsretained}
\begin{document}

\copyrightyear{2020}
\acmYear{2020}
\acmConference[MM '20]{Proceedings of the 28th ACM International Conference on Multimedia}{October 12--16, 2020}{Seattle, WA, USA}
\acmBooktitle{Proceedings of the 28th ACM International Conference on Multimedia (MM '20), October 12--16, 2020, Seattle, WA, USA}
\acmDOI{10.1145/3394171.3414691}
\acmISBN{978-1-4503-7988-5/20/10}

\title{Personal Food Model}


\author{Ali Rostami}
\affiliation{%
  \institution{University of California Irvine}
  \streetaddress{Donald Bren Hall, 3209}
  \city{Irvine}
  \country{United States of America}}
\email{rostami1@uci.edu}

\author{Vaibhav Pandey}
\affiliation{%
  \institution{University of California Irvine}
  \streetaddress{Donald Bren Hall, 3209}
  \city{Irvine}
  \country{United States of America}}
\email{vaibhap1@uci.edu}

\author{Nitish Nag}
\affiliation{%
  \institution{University of California Irvine}
  \streetaddress{Donald Bren Hall, 3209}
  \city{Irvine}
  \country{United States of America}}
\email{nagn@uci.edu}

\author{Vesper Wang}
\affiliation{%
  \institution{University of California Irvine}
  \streetaddress{Donald Bren Hall, 3209}
  \city{Irvine}
  \country{United States of America}}
\email{tianjw1@uci.edu}

\author{Ramesh Jain}
\affiliation{%
  \institution{University of California Irvine}
  \streetaddress{Donald Bren Hall, 3209}
  \city{Irvine}
  \country{United States of America}}
\email{jain@ics.uci.edu}
\renewcommand{\shortauthors}{Rostami, et al.}

\begin{abstract}
  Food is central to life.  Food provides us with energy and foundational building blocks for our body and is also a major source of joy and new experiences. A significant part of the overall economy is related to food.  Food science, distribution, processing, and consumption have been addressed by different communities using silos of computational approaches \cite{Min2019AComputing}. In this paper, we adopt a person-centric multimedia and multimodal perspective on food computing and show how multimedia and food computing are synergistic and complementary.
\newline
  Enjoying food is a truly multimedia experience involving sight, taste, smell, and even sound, that can be captured using a multimedia food logger. The biological response to food can be captured using multimodal data streams using available wearable devices. Central to this approach is the Personal Food Model.  Personal Food Model is the digitized representation of the food-related characteristics of an individual. It is designed to be used in food recommendation systems to provide eating-related recommendations that improve the user's quality of life. To model the food-related characteristics of each person, it is essential to capture their food-related enjoyment using a Preferential Personal Food Model and their biological response to food using their Biological Personal Food Model. Inspired by the power of 3-dimensional color models for visual processing, we introduce a 6-dimensional taste-space for capturing culinary characteristics as well as personal preferences. We use event mining approaches to relate food with other life and biological events to build a predictive model that could also be used effectively in emerging food recommendation systems.

\end{abstract}



\maketitle

\section{Introduction}

\begin{center}

"One cannot think well, love well, sleep well, if one has not dined well."  - Virginia Woolf
\end{center}

Food is a significant determinant of human quality of life. Food provides the energy and nutrients essential for health and is a significant source of personal enjoyment and social fabric. In many instances, pleasures of eating conflict with the optimal nutritional needs of the person's physiological well-being, and is the leading cause of the substantial increase in diet-related diseases such as obesity, diabetes, and hypertension \cite{FoodHealth}, \cite{Schulze2018FoodPrevention}.  An important question is: why do people enjoy food \cite{NewtonAndersonEveryoneCulture}? People working on improving the enjoyment aspect of food, particularly chefs and food industry, have primarily ignored the health, and those focused on health (doctors and nutritionists) usually consider the enjoyment aspect secondary \cite{McclementsFUTUREEAT}, \cite{Kale2020TracingEntries}. This disconnect in the two approaches has led to the current situation with the widespread increase in food-related illnesses. An important fact is:  what I like to eat is not necessarily what my body likes \cite{Mai2011ImplicitResearch}, \cite{Kale2020TracingEntries}.  Can we satisfy both me and my body?

Food and nutrition have their roots in multimedia and multimodal elements \cite{Spence2015MultisensoryPerception}. Food experience requires the participation of audio, visual, tactile, gustatory, and olfactory senses, and prior experiences play a crucial role \cite{Sarabian2017AvoidanceChimpanzees}.  The food we perceptually enjoy is a complete multimedia experience \cite{McclementsFUTUREEAT}, \cite{Spence2015MultisensoryPerception}, which extends further to an extensive multimodal effect in the body, impacting the physiology and biochemistry of the individual. A multitude of sensors can measure the relationship between foods and the individual through the dynamic health state variables \cite{Nag2018Cross-modalEstimation}. These include readily available sensors that provide continuous data collection for blood glucose, heart rate, perspiration rate, and body temperature \cite{KasaeyanNaeini2019AnMonitoring}.

Food is a multimodal experience that enriches personal life and enhances social rituals important to humans. However, we have not studied all aspects of food in a unified computational framework like many other aspects of life, such as social networks, sports, and entertainment. Recently Min et al. \cite{Min2019AComputing}  put together a computational framework around different silos of food. They adopt a diverse data-centric perspective and define food computing as, \emph{computational approaches for acquiring and analyzing heterogeneous food data from disparate sources for perception, recognition, retrieval, recommendation, and monitoring of food to address food-related issues in health, biology, gastronomy, and agronomy.} 
This exhaustive and inclusive approach to food computing will help understand different aspects of the food ecosystem and how they impact each other. 

\begin{figure}[!ht]
  \centering
  
  \includegraphics[width=0.8\linewidth]%
    {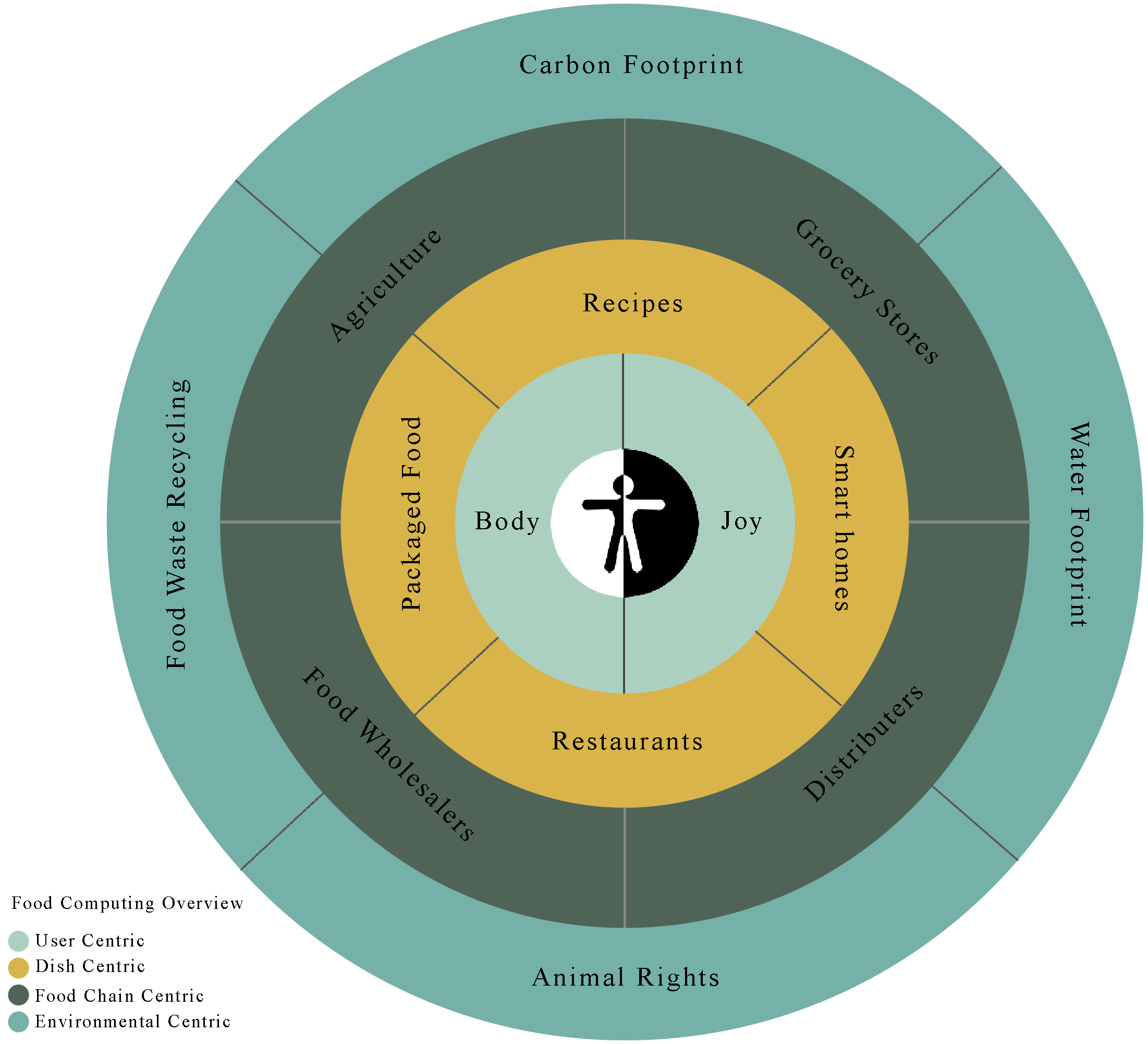}
    \caption{Personal Food Computing Overview}
    \label{fig:personcentric}
\end{figure}

This paper looks at the food ecosystem from a person-centered perspective.  Our goal is to study how food affects a person's life and how the food ecosystem may be affected by choices made by people, as shown in Figure \ref{fig:personcentric}. 

Food serves two crucial but closely related functions of maintaining biological health state and personal enjoyment in life. 
Food items, listed in the dish-centric layer, meet the personal food needs of individuals.
A group of food producers and distributors are part of the next layer that we show as the food chain.  Finally, each item produced, distributed, served, and consumed has a specific effect on the environment.  

In this paper, we present a computational framework for building a Personal Food Model (PFM) that is essential to help people identify the right food, at the right place, in the right situation, at the right price. PFM is an essential component of emerging food recommendation systems to address challenges in different aspects of businesses as well as individuals' health \cite{Min2019FoodChallenges}. We consider the aspects of food that satisfy the two crucial needs of a person: enjoyment and sustenance. Different groups of people have studied these two aspects. We believe that there is an excellent opportunity to bring these disjoint areas together using a computational framework centered around multimedia.

The most important contribution of this paper is the unified personal model of culinary multimedia experience and biological health aspects. We use this model in a complex recommendation system that considers food items as a combination of features contributing to both enjoyment and sustenance and optimizes specific health outcomes such as sleep quality. 
We model a person by analyzing their multimodal food experiences as well as complex contextual factors related to different food items and dishes. The recommendation system then tries to optimize factors related to both enjoyment and sustenance by selecting correct food dishes in a given context.

We present the personal food model (PFM), as a critical, relevant, and timely challenge for multimedia and multimodal research.  We present these ideas by 

\begin{figure*}[h]
  \includegraphics[width=0.8\textwidth]{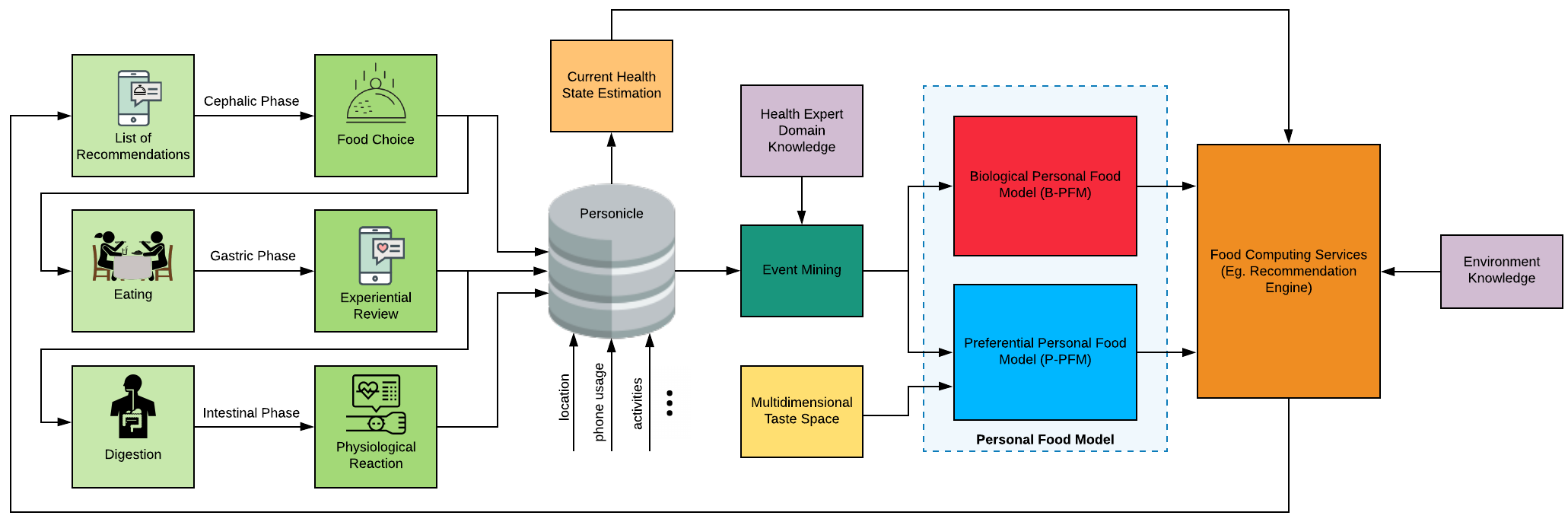}
  \caption{Food Recommendation Architecture: Data from the 3 digestion phases are being collected alongside other data-streams to create the PFM: the heart of Food Recommendation.}
  \label{fig:PFMSystemsOverview}
\end{figure*}
\begin{enumerate}
\item Reviewing existing work in multimedia that peripherally touched food computing but did not address real challenges. We believe this was due to the absence of a clear challenge and application. We show that characteristics of food items and the food preferences of a person can be understood by combining visual, olfactory, culinary, and tactile (texture) aspects of food and eating environment.
\item Discussing essential aspects of personal food computing that will benefit significantly from multimedia technology and offer new challenges for the multimedia community. Notably, we discuss a multimodal food logging platform for building PFM and using it in a novel food recommendation platform. This may open a prominent application area for multimedia computing.
\item Presenting early components of personal food model based on multimedia computing, but require significant new research to create applications that may rival any past multimedia applications.
\end{enumerate}

As discussed in subsequent sections, these are primarily multimedia challenges that will open new paradigms in multimedia computing and communications and will help people enjoy good food and be healthy.

\section{Personal Food Model}

PFM is the digitized representation of the food-related characteristics of an individual. It can be used in food recommendation systems to provide eating-related recommendations that improve the user's quality of life. Many factors affect and limit a simple eating decision. However, this problem has not been modeled in a comprehensive framework to study food as a multimedia experience, including taste, visual, social, and experiential factors. We show how PFM can predict the user's multimodal food preferences in different contexts. We accomplish this using different data streams captured from the user, such as location history \cite{Nag2019SynchronizingMonitoring}, vital sign streams, and food intake logged using text voice and photos. In future works, we plan to expand the sources of information we use to create the personal model and focus on using many other data streams such as the user's calendar, social media, and transaction history.

PFM encompasses a complex nature as it contains many dimensions. The Biological part captures how Food items can satisfy nutritional needs for certain goals such as weight loss or improved performance in athletics \cite{Nag2019ALife}. Furthermore, Contextual understanding of the user needs must be layered for best computing real-time needs \cite{Nag2017LiveEngine}. Other biological and life events may also impact the food events indirectly and needs to be added to the model \cite{Pandey2020ContinuousRetrieval}.

Figure \ref{fig:PFMSystemsOverview} shows how the personicle collects different data streams over a long period \cite{Jal2014Personicle:Events}. Events from the personicle are fed to the PFM which consists of two parts. We define the Biological PFM of the user to capture the body's reactions to different food items including allergic reactions and nutritional needs.
The biological model is an important factor in each food decision we make, but it is not the only factor.  We also create the user's taste profile, which constitutes the Preferential Personal Food model for the user. User's taste profile contains the information about the food items which the user has experienced in the past, and it may also reveal dishes that the user has never tried.
\begin{figure}[!ht]
  \centering
  
  \includegraphics[width=0.7\linewidth]%
    {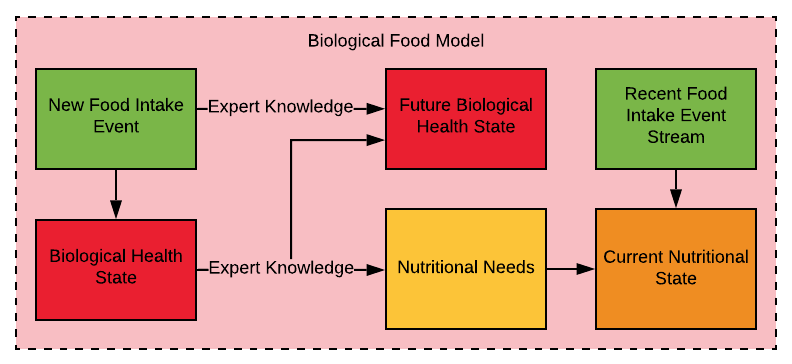}
    \caption{The interactions in the biological food model}
    \label{fig:biological-model}
\end{figure}
\subsection{Biological Personal Food Model}
The Biological Personal Food Model (B-PFM) must consider how food is related to the health state of the individual \cite{Nag2020HealthEstimation}. This model should also extend to how the user may want to change their health state towards a specific goal \cite{Nag2019ALife}. The B-PFM focuses on the user's dynamic health and nutritional needs \cite{Nag2017LiveEngine}, \cite{Nag2017PocketLocation}.

Building the Biological PFM in a purely data-driven manner is a daunting task.
Even though some apps like MyFitnessPal collect food intake and activity data from the user, they only focus on a limited fitness aspect and cannot be extended to a general biological model. However, instead of finding the patterns solely based on user data, we propose a hybrid approach using patterns obtained from domain knowledge to form a rule-based population model. We personalize this model as we collect more data. These rules capture the impact of food on biological parameters.
For example, research shows that eating heavy meals before bedtime could lower the quality of sleep. We collect a selection of such sequences from expert domain knowledge and calculate the probability of validity for each of these patterns in different contexts. This set of context-driven rules form the B-PFM. 

We also need to understand how food items and food events impact different aspects of the health state of the individual \cite{Nag2020HealthEstimation}, \cite{Nag2017PocketLocation}. The user could have multiple health goals that might lead to conflicting recommendations (eg. diet for weight loss and sleep improvement). Therefore, it is important to keep the balance between different biological goals while also including static personal factors such as allergies, intolerances, and genetic factors in this computation.
Figure \ref{fig:biological-model} shows how the current nutritional state is impacted by the food intake based on the particular needs of the user at the current biological state. In the event mining section, we describe how we turn the expert knowledge into active rules and validate them based on the user's data to predict the future biological health state.

\begin{figure}[!ht]
  \centering
  
  \includegraphics[width=1.03\linewidth]%
    {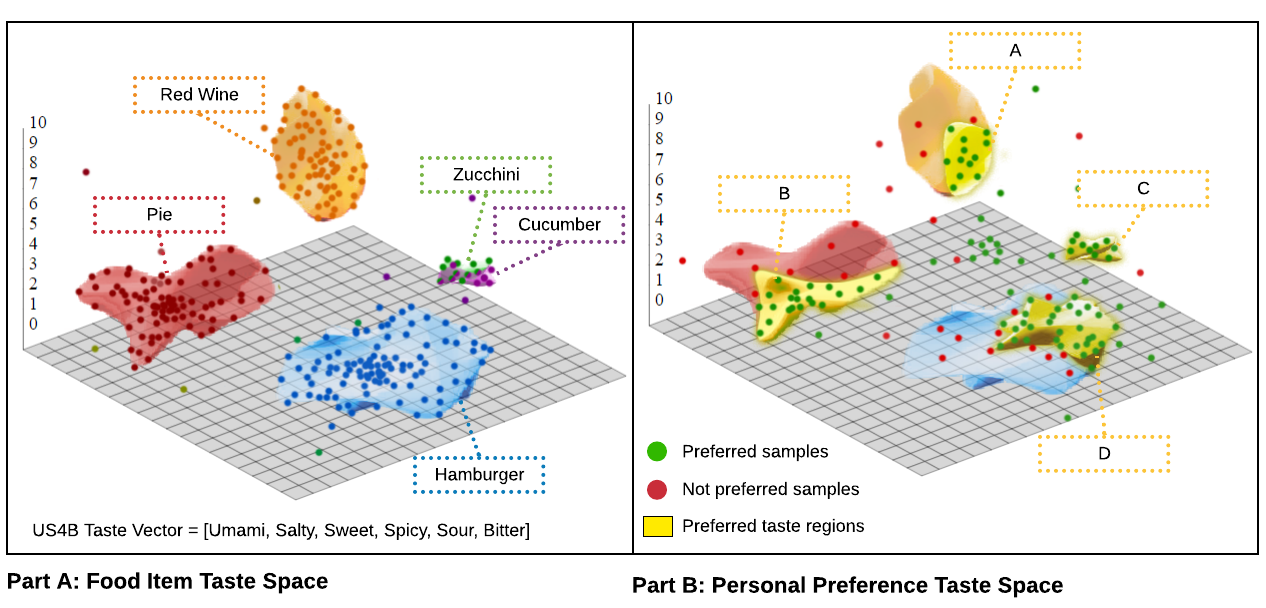}
    \caption{Visualization of the US4B Taste Space. Part A: The collection of all taste samples from all users determine the hypervolume of the taste range of food items. Part B: Past taste sample values and  ratings from the user determine user's preferred taste region within the food item taste range hypervolume.}
    \label{fig:hd-taste-space}
\end{figure}
\subsection{Preferential Personal Food Model}
We propose a novel approach to quantify and describe human taste perception to create the Preferential Personal Food Model (P-PFM). Based on the current state of the art methods food recommendations either ignore the preference model completely and just focus on healthy recommendations \cite{Rehman2017Diet-right:System}, or try to find the preferred ingredients of the user by asking the user to rate a long list of ingredients and dishes without really understanding why the user likes an ingredient \cite{ElahiInteractionSystem}.
We introduce a taste space using six taste primaries called the US4B taste model. The US4B taste model is a multidimensional additive taste model, in which umami, salty, sweet, spicy, sour, and bitter taste (USSSSB) are added together in various ways to reproduce a broad array of tastes.
The RGB color space has been the foundation of many advances in multimedia technology such as digital displays, virtual reality and 3D printing. The US4B taste space can be the key to future food-related technologies that were not possible without this foundation. 

Each food item will have an exact value in the US4B channels which determine its taste. As Figure \ref{fig:hd-taste-space} shows, we create a Hyperdimensional Taste Space (HD-Taste-Space) and calculate a region for each food item. An unripe mango from Brazil is going to have a different vector value in the HD-Taste-Space compared to a ripe mango from India whereas zucchini and cucumber samples share the same taste region. Therefore by sampling different instances we associate a hypervolume in the US4B Taste Space to each food item which is shown in Figure \ref{fig:hd-taste-space} part A. Then we use the recipe databases to estimate the hypervolume containing the possible tastes for the dish in the hyperdimensional space. The state of the art finds correlations among recipes based on their ingredients \cite{Kuo2012IntelligentIngredients} but there has been no research to really understand the taste of the dish based on the recipe as a multidimensional media. To create the P-PFM we map the food log to the HD-Taste-Space to compute a hypervolume representing the user's preferred taste regions. The user's preferred taste regions in the US4B taste space is the most important part of the P-PFM. It contains the information about the food that the user likes and has experienced before, and can also predict the food that the user has never tried but might like because it lies within the user's region of interest. Knowing the preferred regions, we can search for healthier food items within the user's preferred range of taste. Diet soda is a classic example of this concept. It has similar taste, texture, smell and visual cues compared to a normal soda but it has different effects on the biology. Having the food items in the US4B taste space and finding the user's preferred taste regions in this space enables us to come up with better food options tailored for each individual's specific needs and taste preference.

\section{Collecting Data: Food Chronicle}

Models are built using data. Most successful search engines, social media, and e-commerce systems utilize personal models to provide people with the right information, at the right time, in the right context, usually even before a user articulates his need \cite{AdomaviciusTowardExtensions}. Personal food model plays the same role in food recommendation systems \cite{Min2019FoodChallenges}. We need to log food consumed by a person and all the relevant metadata over a long period for the user. While initial food logging efforts required cumbersome manual food diaries, smartphones and cameras can drastically improve the quality and ease of logging. Aizawa \cite{FoodLog:Applications} was the first multimedia researcher to champion the idea of logging food using a smartphone camera and remains a very active researcher. Applications for camera-based food-logging have been developed in many other countries \cite{Chen2016Deep-basedRetrieval}, \cite{Bossard2014Food-101Forests}. 
Multimedia and computer vision research communities have been actively exploring food-logging systems. These systems use computer vision techniques to recognize items, their ingredients, and even the volume consumed by the user \cite{Chen2016Deep-basedRetrieval}, \cite{Oh2018MultimodalJournaling}. Unfortunately, there is no generalized logger for international food, and identifying ingredients and volume remains a challenge. A useful review of many visual approaches and descriptions of databases used for training is included in \cite{Min2019AComputing}.

Conversational voice interfaces are becoming quite popular, making rapid progress.  Systems like Alexa and Siri are available at home, in phones, and watches. People can report what they eat, volume, and reaction to food using a simple sentence. Many packaged food and processed ready to prepare food items have barcodes.  Since barcode readers are now omnipresent even in smartphones, one can get all food information from these.  Some sensors measure muscle activity and try to infer food items from that.  These are placed on the chest, near the ear, or neck \cite{Chu2019RespirationSensors}.  These have shown some progress in recognizing eating events but have not gone much beyond that yet.

\begin{figure}[!ht]
  \centering
  
  \includegraphics[width=\linewidth]%
    {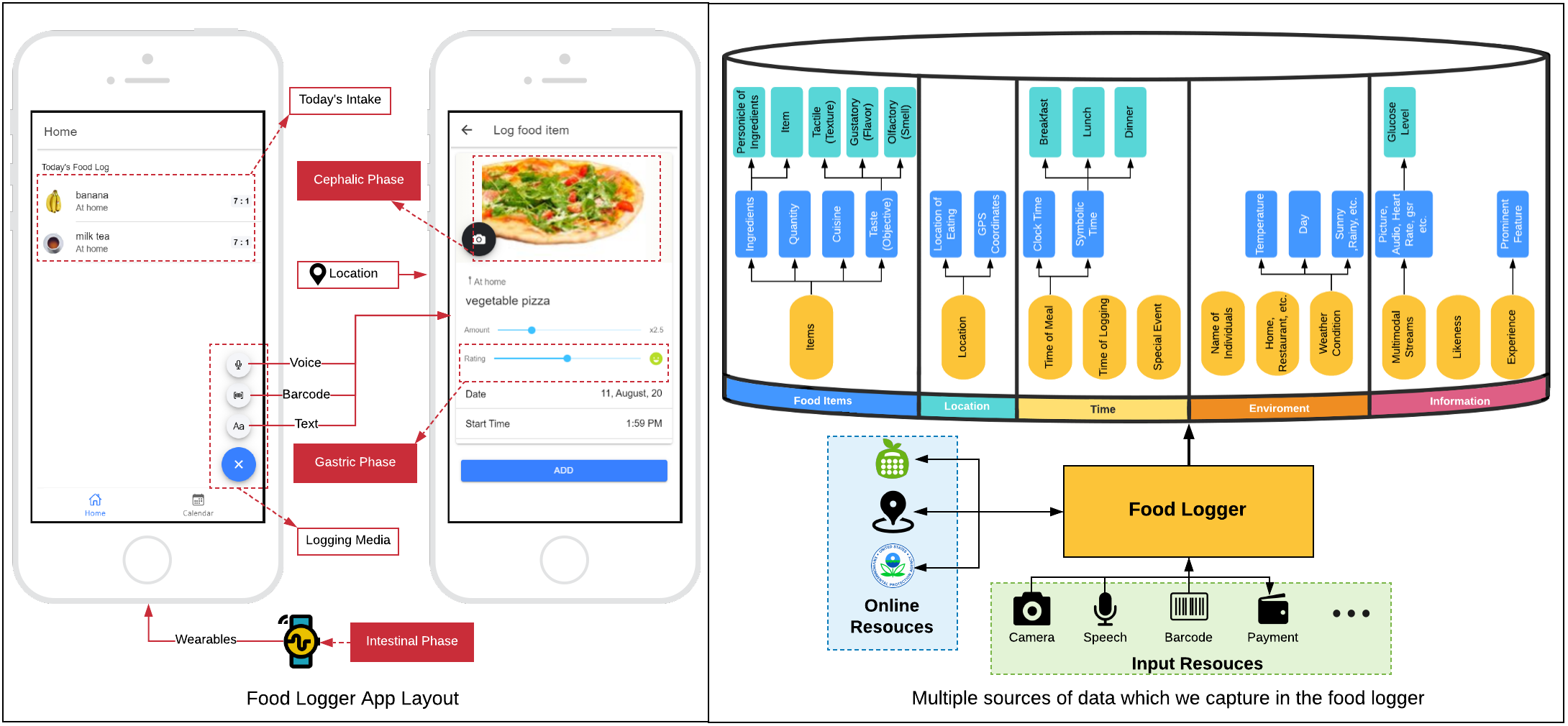}
    \caption{Food logging will use multimedia input sources and complement information from online databases to log each meal and all metadata related to the meal. It captures information about the food (dish name, ingredients, quantity), location (place of eating), time (eating and logging), social context (companions), causal aspects (nutritional and flavor information), and multimedia and experiential information about the food.}
    \label{fig:foodlogger}
\end{figure}

We propose a multimedia food logging platform, shown in Figure \ref{fig:foodlogger}, that could use many relevant sources to log food items and find all information that may be needed to build a PFM. Multimedia uses complementary and correlated information and provides more comprehensive and precise information than any one medium. Moreover, we will keep adding new sensors and technologies to keep the logger useful.

This is the beginning of building towards a robust multimedia solution to the problem of logging.  There are three important aspects to this platform:
\begin{itemize}
    \item We can design a multimedia platform that uses visual food recognition, speech-based systems, payment based options, barcodes, sensors to determine food chewing and content of the food, and several similar emerging approaches.
    \item Once a dish or food item is recognized and the amount consumed is known, systems must find the nutritional data using governmental or commercial databases.  Similarly, weather information, social context, and other metadata related to food required by the PFM may come from other sources. 
    \item The log must contain the user's reaction both in terms of enjoyment and bodily reaction.  The enjoyment information may come from asking the user, and the bodily reactions may come from sensors such as heart rate, glucose measurement, and respiration rate.
\end{itemize}

\subsection{Utilizing Other Knowledge and Data sources}
We enrich the food events with associated nutritional, culinary, and contextual information using databases from different public and private organizations. These include nutrition (NutritionIX, USDA food database), weather, air-quality (airnow provided by EPA), and place (Google Places, Yelp).
\newline
We may also want to capture some biomarkers characterizing the health of the person. These parameters may be continuously recorded and could be used to identify physiological responses to food items \cite{Oh2018MultimodalJournaling}. A personicle like system \cite{Oh2017FromChronicles} can capture this information, and the time-indexed nature of the data and events makes it readily available for associating and analyzing with the food events.

\subsection{Data Model for the Foodlog}
Food logs are collected for 
\begin{itemize}
    \item Building PFM to understand the nutritional requirements and taste preferences of the user.
    \item Understanding the health state of the user.
\end{itemize}

These two goals may require different information from the food log. Building PFM requires as much longitudinal data as available, while health state estimation requires PFM and recent lifestyle and biological data. We need to keep these goals in mind while designing the food log.
We have followed the HW5 (how, what, when, why, where, what) model as described in \cite{XieEventStreams}, \cite{WestermannTowardApplications} to identify what information can fully describe a food event and maximize its utility for a variety of applications. The different aspects and associated information are detailed in figure \ref{fig:foodlogger}.
There can be three types of data collected:
\begin{enumerate}
    \item Observed data: Directly captured using a sensor.
    \item Derived data: We can derive some data and information using sensors and knowledge sources.  This information will depend on the algorithms and data sources used. 
    \item Subjective data: The system may prompt the user or some other human source to get specific information.  This data is prone to errors as it depends on human perception.
\end{enumerate}
We should utilize the different types of measurements in different manners to minimize the error in our analyses and predictions.

\subsection{Current Status}

In this paper, we describe the data and knowledge needed to build a PFM directed at improving sleep quality. We considered that the sleep quality is affected by stress, activity, and food \cite{Azimi2019PersonalizedStudy}.  We are implementing a food logging platform.  We decided to focus on data collection using voice, text, and barcode for the current version.  We will include visual recognition approaches soon.

We add food metadata in the log using weather and reverse-geo databases.  The foodlogger asks the user about their reaction to each item entered.  We used NutrionIX platform to get information about calories and nutrients in each food item.  The current food logger has information about how much a user likes a dish to build the Preferential Personal Food model. However, the information about the taste and flavor of a food item is not readily available from any source.  We are working towards deriving such information about food items from different sources.  This is an excellent open opportunity for the multimedia community to take the lead in solving this critical problem.
\begin{figure}[!ht]
  \centering
  
  \includegraphics[width=0.9\linewidth]%
    {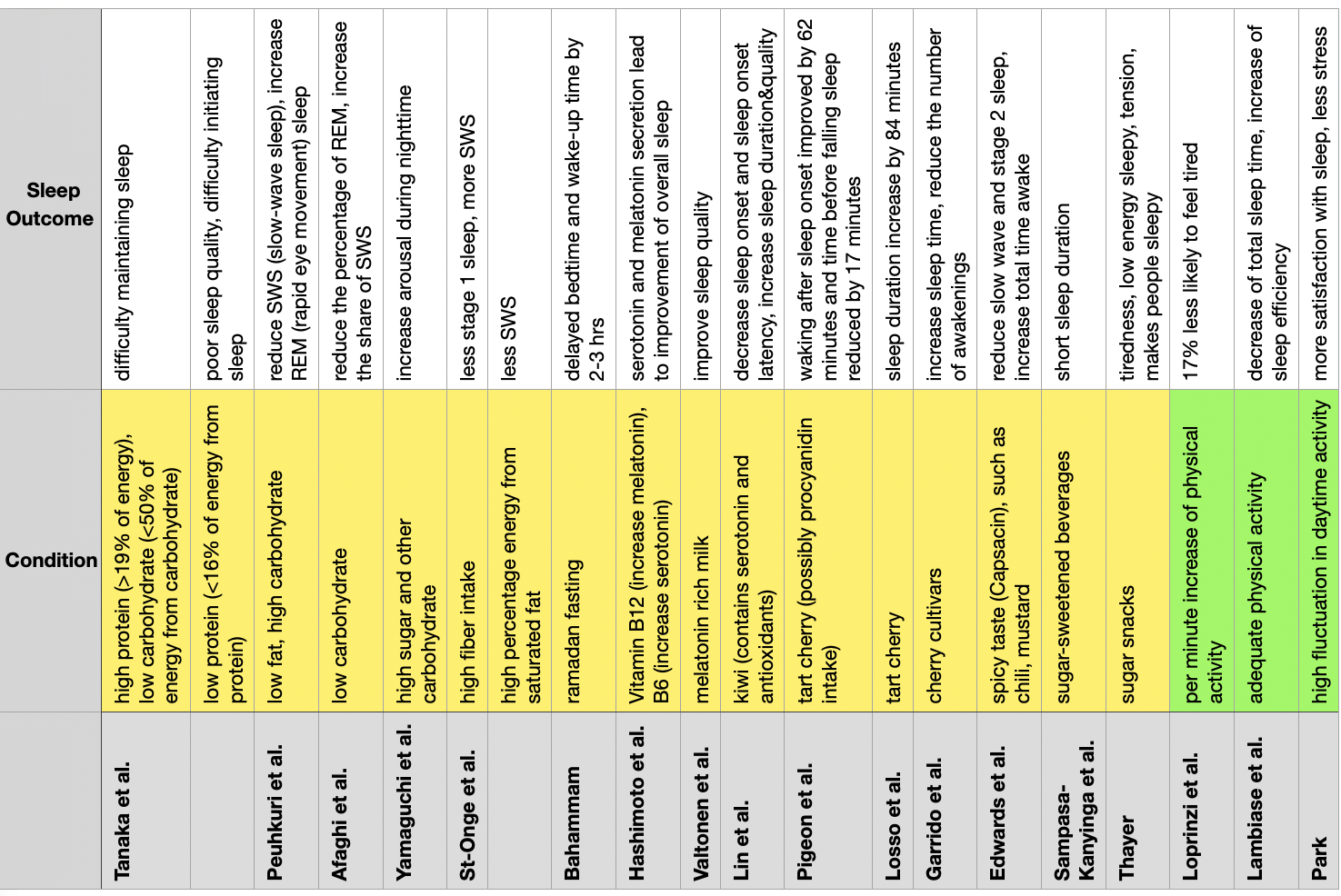}
    \caption{Attributes of food events that impact sleep quality. These relationships form the basis for the Biological Personal Food Model.}
    \label{fig:relatedresearch}
\end{figure}
\section{Knowledge Collection}

As stated in the previous sections, the personal model should be able to incorporate existing knowledge sources. We have surveyed papers that explore the relationship between dietary inputs and sleep outcomes.
We summarize our findings in figure \ref{fig:relatedresearch}.
We found that macro nutrients have a great impact on sleep outcomes \cite{Tanaka2013AssociationsWorkers}, \cite{Peuhkuri2012DietaryMelatonin}, \cite{Peuhkuri2012DietQuality}, \cite{Yamaguchi2013RelationshipRegularity}, \cite{Afaghi2008AcuteIndices}, \cite{St-Onge2016FiberSleep}. Some micronutrients contribute to melatonin secretion, and hence can have significant impact on sleep quality \cite{Hashimoto1996VitaminHumans}, \cite{Valtonen2005EffectsSubjects}. Additionally, there are some studies that explore the effect of specific food items such as kiwi fruit \cite{Lin2011EffectProblems} and cherries \cite{Pigeon2010EffectsStudy}, \cite{Losso2018PilotMechanisms}, \cite{Garrido2013AAging.} on sleep. Some chemicals responsible for specific taste such as capsaicin \cite{Edwards1992SpicyThermoregulation} and sugar \cite{Sampasa-Kanyinga2018SleepAdolescents}, \cite{ThayerEnergyExercise} can also impact sleep. Fasting contributes to the change of bedtime \cite{BaHammam2013TheAssessment} as well.
\newline
We have also included some studies about the impact of exercise and physical activity on sleep \cite{Loprinzi2011Association2005-2006}, \cite{Loprinzi2012TheWomen}, \cite{Lambiase2013TemporalWomen}, \cite{Park2014AssociationsAdolescents} as it is an important confounding variable that impacts both nutritional needs and sleep quality.

\begin{figure}[!ht]
  \includegraphics[width=\linewidth]{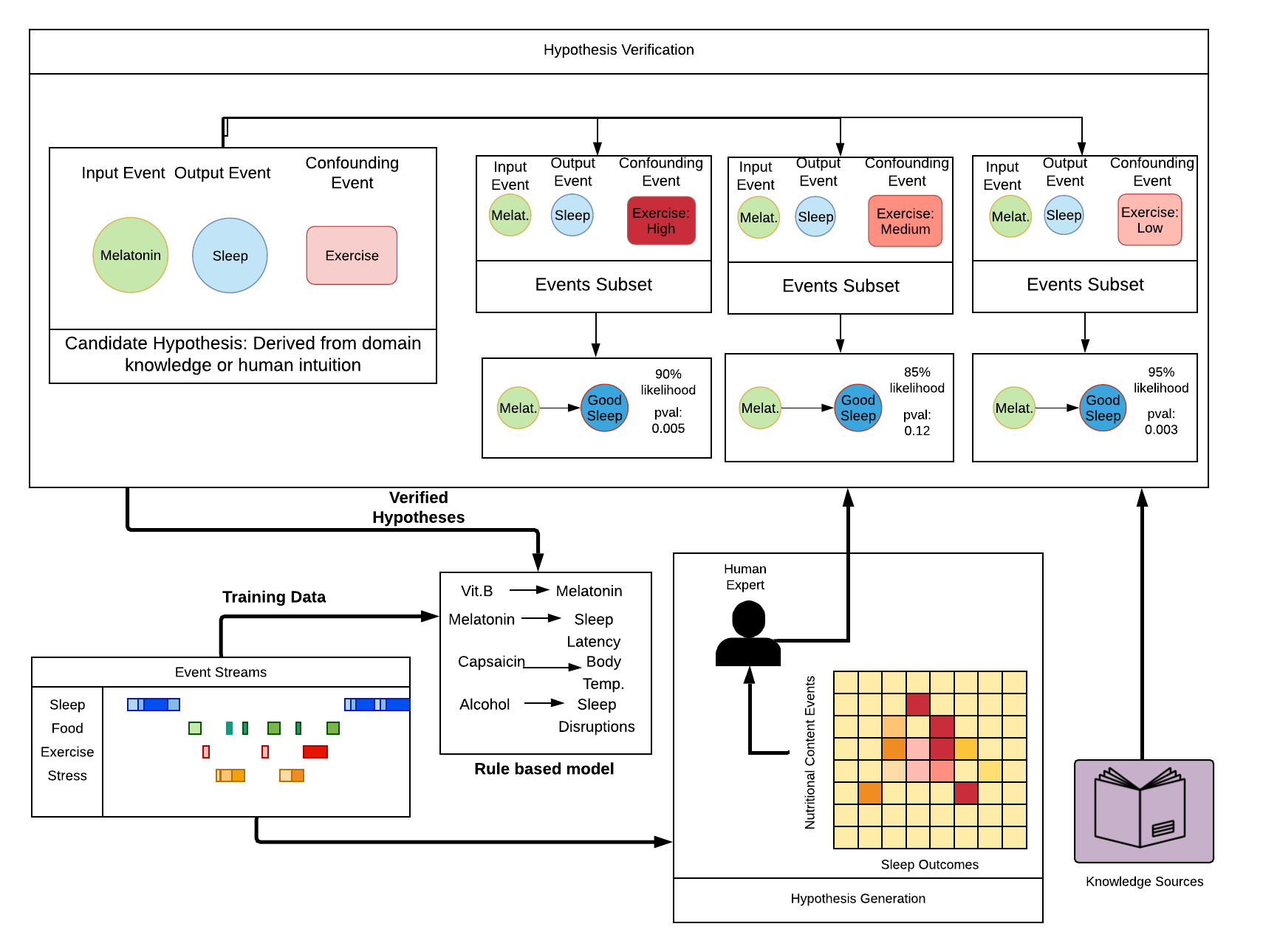}
  \caption{Event Mining workflow: Hypothesis generation operators allows us to find frequently occurring sequences of events. These can be converted to hypotheses by including confounding variables and can then be tested in presence of these confounding factors using hypothesis verification operator. These verified hypotheses serve as a rule-based model for the user's behavior.}
  \label{fig:SleepModel}
\end{figure}

\section{Event Mining}
Multimedia research in event mining has focused on event recognition and situation understanding (eg., sports and surveillance videos). There has not been much research on how we can utilize event mining to run n-of-1 experiments using a person's events and data streams and derive rules that describe their behavior in different situations.
Event mining allows us to find patterns and relationships between different events in our daily lives. We can find relationships between different events in a person's lifelog data and derive an explainable personal model \cite{PandeyUbiquitousHealth}. 

Event mining results in rules of the form $Event_i \xrightarrow{C}Event_o$, where $Event_i$ is the input event, and we want to find out its effect in the occurrences of the outcome, $Event_o$. $C$ defines the set of confounding variables and temporal conditions that might affect this relationship. For Biological-PFM, the input events are lifestyle events that have a causal impact on some observable biological outcome \cite{Pandey2020ContinuousRetrieval}. While, in the Preferential-PFM, the input events capture the contextual situations that affect the user's culinary preferences.
This view of events and their impacts is in line with the potential outcomes framework for causal inference (provided the required assumptions, eg., SUTVA are valid) \cite{Rubin2005CausalDecisions} and are explored in detail later in this section.
\newline
We perform a two-step analysis with a human expert acting as an intermediary to select non-spurious relationships. The event patterns language described in \cite{Jalali2016InteractiveStreams} allows us to describe the relationships as temporal patterns of events. \textbf{Hypothesis generation} is used as a preliminary investigation tool that allows a human expert to identify any behavioral patterns in the form of events co-occurrences and \textbf{Hypothesis verification} tells us whether the relationship is causally significant in the presence of the confounding variables.

\subsection{Hypothesis generation: Discovering new behavioral patterns}

Users' event logs contain all of their daily habits and biological responses to different events. Hypothesis generation operators allow us to discover these habits and patterns that can be tested and used for prediction.
This step of the analysis is mostly data-driven and starts with a human expert specifying the event streams that they believe to be correlated. The output is a heat map with different combinations of events occupying different positions (Figure \ref{fig:SleepModel}).
The patterns with relatively higher frequency may represent a significant relationship and selected for hypothesis verification.
\newline
The frequent patterns would then need to be converted to candidate hypotheses. The user would need to specify the cause and effect events along with any confounding factors. This hypothesis can then be verified using the hypothesis verification operator.


\subsection{Hypothesis verification: Verifying patterns under different contexts}
Users can also verify their beliefs by encoding those as patterns of events and specifying the variables that define the contextual situation. 
We defined these patterns using scientific literature, as described in the previous section.
Each occurrence of the pattern represents an instance of the input event (treatment), and we want to measure its impact on the outcome. Thus, each occurrence of the pattern becomes a single unit in the potential outcomes framework \cite{Rubin2005CausalDecisions}, and we can compare different units while matching them by the confounding factors, to estimate the causal effect of the treatment.
\newline
Once we have found all the pattern occurrences and the confounding variables, we follow a two-step process to find the validity of the rule. 

\begin{enumerate}
    \item \textbf{Find similar situations based on confounding variables (Contextual Matching)}. The confounding variables define the situation in which the input event (treatment) occurs, and can affect the event relationship we want to analyze. Therefore, we want to compare the events that occur in similar contexts and compare the impact of the input event on the outcome in an unbiased manner. We can do it by either clustering the values of confounding variables or converting the confounding variables to events and find matching confounding event patterns. 
    \item \textbf{Find the validity of the relationship for each situation.} Once we have performed the contextual matching, for each contextual group, we can find the effect of the treatment on the outcome using an appropriate statistical test. We can compare the difference in the outcome for different input events, and this would tell us the relative causal effect of the different input events. 
\end{enumerate}
This two-step hypothesis verification allows us to simulate an N-of-1 experiment on the user's event log while also incorporating the existing scientific knowledge in the form of candidate hypotheses and identifying confounding variables. 

\subsection{Deriving Personal Food Model using Event mining}
We need to analyze the food log in conjunction with other events from the personicle to create an explainable and personalized food model for every individual. The model would predict the impact of food events on other aspects of a person's life; and how different lifestyle and biological factors impact our food choices. In this paper, we are exploring the relationship between food events and sleep outcomes; therefore, we will include behavioral factors that would impact these two events, such as physical activity (exercise, step count).

We can identify different behavioral habits of the user using hypothesis generation operators. We can also visualize the relationship between various nutritional factors and different sleep outcomes to find if these are worth exploring further. Once we have identified such relationships, we can start verifying these hypotheses.
We derive the hypotheses from data or existing biomedical literature. These relationships have been detailed in the previous section and are also depicted in figure \ref{fig:relatedresearch}.
\newline
Figure \ref{fig:SleepModel} shows the complete event mining process for the personal food model. The verified hypothesis contain event relationships that hold true in the specified contextual situation.
These relationships form a set of rules with varying degrees of accuracy in different contextual situations. For example, if we have verified that cow's milk has a positive impact on sleep latency, then the relationship would be quantified in the form of minutes reduced in latency. It will have a different value for different contextual situations described by physical activity, day's meals, and last night's sleep. These rules could thus be used to identify the potential outcome of different foods and recommend items with the desired sleep outcome.

\section{Going Forward: Multimedia for Personal Models}

Though this paper focuses primarily on personal food models, it is really about building personal health models using disparate data and information sources.  A personal health model's importance is apparent in these days of a pandemic that has disrupted lives globally.  In this section, we discuss interesting challenges that we need to address.  We believe that multimedia computing offers concepts, techniques, and practical experiences related to key areas mentioned in the paper.

\begin{enumerate}
    \item User Privacy: User privacy and data protection are integral to developing a multimedia personal model. Without adequate security measures the model is unlikely to be widely adopted, regardless of the performance or utility. 
    This is an important challenge for multimedia, artificial intelligence, and privacy and security research groups and we are actively looking for collaborations in this area. 
    There are learning techniques such as federated learning \cite{Yang2019FederatedApplications} that allow us to build models and share insights without taking users' data from their device. We need to incorporate such methods in our platforms so that the users have complete ownership of their data. 
    \item Taste Space: Taste and flavor of food are very complex.  Food taste space depends on the ingredients and recipe as well as visual presentations.  On the other hand, each person has their own preferred taste space that must be determined by observations over a long time. We are exploring 6-dimensional taste space. This is less than the tip of the proverbial iceberg.  Such representations will result in labeling food items better so that people can select what they will enjoy eating and will be healthy.
    \item Multimedia Logging Platform: Multimedia community has focused on food logging using only visual recognition approaches and has been limited only to dish and ingredient recognition.  Food logging is not just recognizing dishes from pictures, but identifying all characteristics of an eating event.  We need to build a multimedia logging platform to collect all food-related information relevant to building PFM. Such logs could be used for studying population for health as well as for business reasons.
    \item Multimodal event detection: The health state of a person is usually estimated by combining multimedia (audio-visual) and multimodal (heart rate, EEG, respiration rate, Glucose content) signals.  Estimation of health state is a great challenge for researchers that will also help ALL humans. 
    \item Multimodal Knowledge Collection: Much of the diagnosis and prescription related to health is multimodal and will require extending traditional knowledge graph \cite{Zulaika2018EnhancingGraphs}\cite{HaussmannFoodKG:Recommendation} techniques.
    \item Event Mining: Mining multiple sequences of event streams detected from disparate data streams is essential for both building models such as PFM as well as for health state estimation.  Event mining may offer more challenging problems in predictive and preventive approaches in several application areas, including health using novel forms of machine learning than object recognition offered in computer vision.  We have already started building a platform for this.
    \item Recommendation System to motivate behavioral change: In context of eating habits, a recommendation system which always promotes the healthiest option is not necessarily the best one. A good recommendation must consider personal food preferences and healthiness together to suggest not just healthy but correct amount of 'healthy and tasty' food. Correct recommendation should be given at the correct place and the appropriate time to motivate behavioral change \cite{Patel2015WearableChange}, \cite{Motivate:Publication}. PFM is the first step towards context-aware recommendation in food domain but this is just the beginning of a long journey.
\end{enumerate}

\bibliographystyle{ACM-Reference-Format}
\bibliography{references}


\end{document}